\documentclass{article}
\usepackage[utf8]{inputenc}
\input{Qcircuit}
\usepackage{algpseudocode}
\usepackage{glossaries}

\usepackage{lscape}
\usepackage{longtable}
\usepackage{amsthm}

\theoremstyle{definition}
\newtheorem{dfn}{Definition}[section]
\newtheorem{exmp}{Example}[section]
\theoremstyle{remark}

\theoremstyle{plain}

\newtheoremstyle{alg}{12pt}{12pt}{\itshape}%
{}{\sffamily}{:}{\newline}{}

\usepackage[section]{placeins}
\begin{document}

\title{New Design of Reversible Full Adder/Subtractor using $R$ gate}

\author{Rasha Montaser \\ Zewail City of Science and
Technology, \\University of Science and Technology, Cairo, Egypt\\ Department of Mathematics and Computer Science, \\Faculty of Science, Alexandria University, Egypt\\
$^\dag$rashamontaser@gmail.com
\\
\and Ahmed Younes$^\S$ \\Department of Mathematics and Computer Science, \\Faculty of Science, Alexandria University, Egypt \\ School
of Computer Science, University of Birmingham, \\Birmingham, B15
2TT, United Kingdom\\
$^\S$ayounes@alexu.edu.eg
\\
\and Mahmoud Abdel-Aty$^\star$ \\ \\ Zewail City of Science and
Technology, \\University of Science and Technology, Cairo, Egypt\\ Department of Mathematics, Faculty of Science, \\Sohag University, Egypt\\
$^\star$mabdelaty@zewailcity.edu.eg}

\maketitle

\begin{abstract}
Quantum computers require quantum processors. An important part of the processor of any computer is the arithmetic unit, which performs binary addition, subtraction, division and multiplication, however multiplication can be performed using repeated addition, while division can be performed using repeated subtraction. In this paper we present two designs using the reversible $R^3$ gate to perform the quantum half adder/ subtractor and the quantum full adder/subtractor. The proposed half adder/subtractor design can be used to perform different logical operations, such as $AND$, $XOR$, $NAND$, $XNOR$, $NOT$ and copy of basis. The proposed design is compared with the other previous designs in terms of the number of gates used, the number of constant bits, the garbage bits, the quantum cost and the delay. The proposed designs are implemented and tested using GAP software.\\

keywords:reversible gates; quantum processors; arithmetic unit; reversible adder; reversible subtractor.
\end{abstract}



\section{Introduction}
Quantum computers offer essential speedup over classical computers. Many researchers believe that, no amount of progress in classical computers, could overcome the power of a quantum computer \cite{citation42}. More computer resources are required to solve computational problems; some problems are impossible to solve using classical computers, because they require ridiculous resources to solve the realistic case of the problem \cite{citation12}. Quantum computers can efficiently solve computational problems, that have no efficient solution on classical computer, Since Classical computers are built using irreversible gates \cite{citation13n,citation42}. 
The main problem in using irreversible gates is the loss of data due to the problem of heat dissipation, since the amount of energy dissipated is proportional to the number of bits erased. To overcome this problem, reversible gates are used, as they do not cause heat dissipation, and thus there will be no loss of data \cite{citation01n,citation02n}.\\

Reversible gates are the building block of reversible circuits. A reversible circuit maps each input vector to a unique output vector and vice versa \cite{citation13n}. Reversible circuits have many applications such as: building quantum computers, bioinformatics, nanotechnology based systems, low power CMOS systems, digital signals processing, optical computation, DNA computing and communications \cite{citation13n,citation13-1n,citation16n}. In reversible computing, the output values are sufficient to recover the input values, the number of inputs must be equal to the number of outputs, each input pattern maps to a unique output pattern. Thus fan-out, feed-back and loops are not allowed. When building a reversible circuit, the following criteria should be considered: minimum number of reversible gates, minimum number of garbage outputs, minimum number of constant inputs and minimum quantum cost. In reversible computers the main building blocks are: Multiplexer, Decoder, Ari thematic logical unit ...etc \cite{citation01n,citation02n,citation04n,citation13n}.\\

Arithmetic adders and subtractors are the fundamental building blocks in many computational units. The expected paradigm shift logic that is compatible with quantum computation requires a compatible implementation of adder and subtractor. They can be used to perform other logical operations, such as multiplication and division \cite{citation04n,citation13n}. Thus many researchers focus on designing adders and subtractors. A method for constructing reversible full adder using two $RG$ gates is introduced in \cite{citation01n}. A reversible full adder using two Peres gates, to minimize the quantum cost is designed in \cite{citation04n}. An efficient full adder circuit based on four Fredkin gates is introduced in \cite{citation07n}. A new reversible gate called $NR$ gate is introduced in \cite{citation11n6}, then used to designed an  optical reversible full adder and subtractor, composed of two NR gates. A new $4 \times4$ reversible gate called $MOG$ gate, which can be used by itself as a full adder/subtractor, is introduced in \cite{citation10n}. The $MOG$ gate is constructed from two $C$ gates, one Fredkin gate and one Peres gate. Two designs using $PRT$ gates are introduced in \cite{citation13n}. These designs are used to build reversible full adder using two $PRT$-2 gates, and reversible full subtractor using two $PRT$-1 gates. A new reversible gate called $TR$ gate is introduced in \cite{citation13-1n}, and then used to build a reversible full subtractor, then the quantum cost of the $TR$ gate is optimized in \cite{citation18n}. The optimized model is used to redesigned the full subtractor. Three designs for one-bit full adder/ subtractor is presented in  \cite{citation14n}, then used to build eight-bit parallel binary adder/subtractor. The first design is composed of five $C$ gates, two Fredkin gates and one $TR$ gate, the second design is composed of two $TR$ gates and two $C$ gates, and the third design is composed of two $C$ gates and two Peres gates.\\       

The aim of this paper is to introduce a new design of half and full Adder/ Subtractor using the $R$ gate introduced in \cite{citation17n}. In this paper, we are going to introduce two designs using the $R^3$ gate, which are: the half adder/subtractor and the full adder/subtractor. Three different $R^3$ gates ($R^3_{3,1,2}$, $R^3_{1,2,3}$ and $R^3_{2,3,1}$) are used in the two proposed designs. This new design is compared with the other full Adder/Subtractor introduced in \cite{citation11n6,citation10n,citation14n}. The new design can be used as low quantum cost reversible ALU, it also has minimum number of garbage bits and minimum number of gates. These designs have advantage over the other designs, that they can be used as adder and/or subtractor using the same gate with minimal cost, in addition the proposed half adder/subtractor design can be used to perform different logical operations, which are: $AND$, $XOR$, $NAND$, $XNOR$, $NOT$ and copy of basis, and the proposed full adder/subtractor design can be used to perform different logical operations, which are: $AND$, $XOR$, $XNOR$ and $NOT$.\\

The organization of this paper is as follows: Section 2 gives a short introduction to the elementary quantum gates, and defines the terminologies used in this paper. Section 3 proposed the new designed half and full adder/subtractor.  The results and discussion are presented in Section 4, where it compares the proposed designs with relevant designs proposed by others. Finally Section 6 concludes the paper.\\

\section{Basic Definitions}
 This section introduces some basic symbols, definitions and terminologies, used in
synthesizing of reversible circuits, to build reversible full adder/subtractor.
 
\begin{dfn}
A Boolean function $f:x \to y$  is said to be reversible,  if
and only if each input vector $x \in X^n $ maps to a unique output
vector $y \in X^n $.  For $n$- inputs/ outputs function, there
are $(2^n)!$ reversible functions, $\forall X  = \{ 0,1\}  $ and $
n \in  Z$ \cite{citation06}.
\end{dfn}

\begin{dfn}
The main reversible gate \({C}^{n } \)\({NOT} \) used to synthesize any reversible circuit, is defined as,

\begin{eqnarray}
\begin{array}{l}
(y_1,y_2, \ldots, y_{n-1};f_{out}) = C^nNOT (x_1,x_2, \ldots, x_{n-1};f_{in}).
 \end{array}
\label{eqn1}
\end{eqnarray}
where $y_i=x_i$ for 1$ \le i \le n-1$ and $f_{out}=f_{in}\oplus
x_1x_2 \ldots x_{n-1}$. $x_1,x_2, \ldots ,x_{n-1}$ are called the control bits
and $f_{in}$ is called the target bit \cite{citation06}.
\end{dfn}

\begin{dfn}
The cost of the circuit is the total cost of all the $n$ gates used to synthesize the circuit
\cite{citation05}.
\end{dfn}

\begin{dfn}
The minimum cost $Minc(g)$ defined for a reversible gate $g$ means that, there exists a realization of $g$ with cost equal to   $Minc(g)$, and there is no any other realization with cost less than $Minc(g)$
\cite{citation05}.
\end{dfn}

\begin{dfn}
Reversible gate $g$ computes a reversible functions $f$ and is bijective
\cite{citation06,citation08}.
\end{dfn}

\begin{dfn}
A permutation  $\sigma {\rm :}A \to A$ is said to be a bijection. It maps an input to an output, from a finite
set \({\rm  A = \{ 1,2,} \ldots {\rm ,}n{\rm \} }\), can be
written as follows,

\begin{eqnarray}
\begin{array}{l}
\sigma {\rm  = }\left( {\begin{array}{*{20}c}
   1  \\
   {\sigma (1)}  \\
\end{array}\begin{array}{*{20}c}
   2  \\
   {\sigma (2)}  \\
\end{array}\begin{array}{*{20}c}
   3  \\
   {\sigma (3)}  \\
\end{array}\begin{array}{*{20}c}
    \ldots   \\
    \ldots   \\
\end{array}\begin{array}{*{20}c}
   n  \\
   {\sigma (n)};  \\
\end{array}} \right)
 \end{array}
\label{eqn3}
\end{eqnarray}
The top row can be eliminated and written as follows,
\begin{eqnarray}
\begin{array}{l}
\sigma {\rm  = }\left( {\begin{array}{*{20}c}
   {\sigma (1)} & {\sigma (2)} & {\sigma (3)} & {\begin{array}{*{20}c}
    \ldots  & {\sigma (n)}  \\
\end{array}}  \\
\end{array}} \right);
 \end{array}
\label{eqn4}
\end{eqnarray}
another notation having a permutation in the form of $\left( {\begin{array}{*{20}c}
   1  \\
  2  \\
\end{array}\begin{array}{*{20}c}
   2  \\
   3  \\
\end{array}\begin{array}{*{20}c}
   3  \\
   1  \\
\end{array}\begin{array}{*{20}c}
   4  \\
   4  \\
\end{array}\begin{array}{*{20}c}
   5  \\
   8  \\
\end{array}\begin{array}{*{20}c}
   6  \\
   5  \\
\end{array}\begin{array}{*{20}c}
   7  \\
   6  \\
\end{array}\begin{array}{*{20}c}
   8  \\
   7  \\
\end{array}} \right)$, it can be written as (1,2,3)(5,8,7,6,5), this notation is called the product of disjoint cycles  \cite{citation06}. 
\end{dfn}

\begin{dfn}
The constant bit is an input to the circuit used to compute some given logical operations
\cite{citation11n6}.
\end{dfn}

\begin{dfn}
Garbage bit is an additional output to the reversible logic gate, that is added to make the number of inputs equal to the number of outputs whenever necessary, in order to achieve reversibility.
\cite{citation11n6}.

\end{dfn}

\begin{dfn}
The quantum cost of any reversible circuit is the number of elementary gates  (2-qubit gates) used to build the circuit \cite{citation05,citation08,citation11}. There are two methods of calculating the quantum cost, which are:   cost015 metric, which consider the quantum cost of any $1\times 1$ reversible gate as $zero$, the quantum cost of any $2 \times 2$ reversible gate as $one$, and the quantum cost of the other reversible gates is calculated by counting the number of elementary gates used to build them
\cite{citation01,citation11n6}. The cost115 metric, is similar to cost015 but it consider the quantum cost of the $1\times 1$ gates as $one$  \cite{citation11n6,citation08,citation11}. In this paper we are using cost015 metric.  
\end{dfn}

\subsection{Reversible Gates}
There are many universal reversible gates, these gates are used to build any reversible circuits, this section introduces some of the popular important gates used to build adder and subtractor circuits.\\

\subsection*{The $NOT$ gate ($N$)}

\begin{figure}
\begin{center}
\[
\Qcircuit @C=0.7em @R=0.5em @!R{
\lstick{x_1}&	&\targ		&\qw			&\rstick{y_1}\\
&&N
}
\]
\end{center}
\caption{The possible $N$ gate over 1-bit reversible circuit.} \label{fig1}
\end{figure}
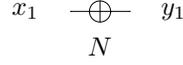

The $N$ gate is a 1-bit reversible gate, it flips the input bits unconditionally. For $1\times 1$ reversible circuit, there is only one $N$ gate with quantum cost equals to $zero$ \cite{citation21}.  Eqn.\ref{eqn5} shows the functionality of the $N$ gate. The circuit representation of the $N$ gate is shown in Fig. \ref{fig1}\cite{citation09}.

\begin{eqnarray}
\begin{array}{l}
N_1 : y_1=x_1\oplus1 = \bar{x}_1, \\
\\
 N_1: (x_1)  \to (1,2), \\
 \end{array}
\label{eqn5}
\end{eqnarray}

\subsection*{The Feynman  gate ($CNOT$)}

\begin{figure}
\begin{center}
\[
\Qcircuit @C=1.5em @R=0.5em @!R{
\lstick{x_1}    &\ctrl{1}                 &\targ            &\qw &\rstick{y_1}\\
\lstick{x_2}      &\targ        &\ctrl{-1}         &\qw &\rstick{y_2}\\
&C_{1,2}  &C_{2,1} 
}
\]
\caption{The possible $C$  gates over 2-bits reversible circuit.} \label{fig2}
\end{center}
\end{figure}
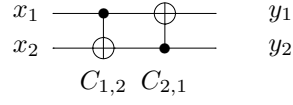

The $CNOT$ gate (also denoted as $gls{C}$ gate) is a 2-bits reversible gate. It takes as input 2-bits, a target bit and a controller bit, then flips the target bit, if the controller bit is set to 1. For $2\times 2$ reversible circuit, there are two $C$ gates, having a quantum cost of $one$ \cite{citation09}.  Eqn.\ref{eqn6} shows the functionality of the $C$ gate. The circuit representation of the $C$ gate is shown in Fig. \ref{fig2}\cite{citation09}.

\begin{eqnarray}
\begin{array}{l}
C^2_{i,j}: \\y_i=x_i,\\ y_j=x_j \oplus x_i,\\
\\
{C}_{{ 1,2}}^{ 2} { :(x}_{ 1} { ,x}_{ 2}  { )} \to {  (3,4)}, \\
{ C}_{{ 2,1}}^{ 2} { :(x}_{ 1} { ,x}{}_{ 2} { )} \to { (2,4)}, \\
\end{array}
\label{eqn6}
\end{eqnarray}
where $i$ and $j \in \{1,2\}$ in any order.

\subsection*{The Square-root $NOT$ gates ($V$ and $V^\dag$)}

\begin{figure}[h]
\Qcircuit @C=1.5em @R=0.5em @!R{
\lstick{x_1}& &\ctrl{1} &\gate{V}  &\ctrl{1} &\gate{V^\dag}  &\qw &\rstick{y_1}\\
\lstick{x_2}& &\gate{V} &\ctrl{-1} &\gate{V^\dag}  &\ctrl{-1} &\qw &\rstick{y_2}\\
&&v_{1,2}&v_{2,1}&u_{1,2}&u_{2,1}
}
\caption{The four possible $v$ and $u$ gates over 2-bits reversible circuit}
\label{fig3-1}      
\end{figure}
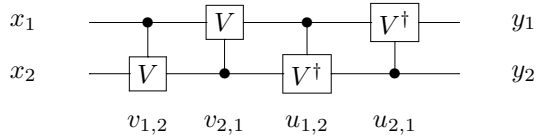

The square-root NOT gate are the
controlled-$V$ ($v$) and the controlled-$V^\dag$ ($u$). They are 2-bits reversible gates, their quantum cost is $one$. For $2\times2$ reversible circuits, there are two $V$ gates and two $V^\dag$ gates,  Fig. \ref{fig3-1} shows their circuit representation. They have the following properties  \cite{citation05,citation09}:\\

\begin{itemize}
  \item$vv=uu=N$,
  \item $uv=vu=I$,
  \item $vN=Nv=u$,
 \item $uN=Nu=v$,
\end{itemize}
where $I$ is the
identity gate \cite{citation05,citation09}. \\

\subsection*{The Toffoli gate $T^3$}

\begin{figure}
\begin{center}
\[
\Qcircuit @C=1.5em @R=0.5em @!R{
\lstick{x_1}         &\ctrl{2}       &\ctrl{1}       &\targ      &\qw     &\qw              &\rstick{y_1}\\
\lstick{x_2}   &\ctrl{1}       &\targ      &\ctrl{-1}      &\qw     &\qw      &\rstick{y_2}\\
\lstick{x_3}   &\targ      &\ctrl{-1}      &\ctrl{-2}      &\qw      &\qw                   &\rstick{y_3}\\
&T_{1,2,3}^3 &\   \ \   T_{1,3,2}^3 &\   \ \ \  T_{2,3,1}^3&
}
\]
\caption{The possible $T^3$ gates over 3-bits reversible circuit.}
\label{fig3}
\end{center}
\end{figure}
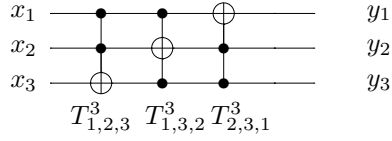

The $T^3$ is a 3-bits reversible gate. It takes as input 2 controller bits and 1 target bit, then flips the target bit, if the control bits are set to 1. For a $3\times 3$ reversible circuit, there are three $T^3$ gates, having a quantum cost equals to $five$. Eqn.\ref{eqn7} shows the functionality of the $T^3$ gate.  The circuit representation of the $T^3$ gate is shown in Fig. \ref{fig3} \cite{citation21}.

\begin{eqnarray}
\begin{array}{l}
T^3_{j,k,l}: \\y_j=x_j,\\ y_k=x_k,\\ y_l=x_l \oplus x_jx_k,\\
\\
 T_{1,2,3}^3 {\rm :(}x_1 ,x_2 ,x_3 {\rm )} \to { (7,8)}, \\
 T_{1,3,2}^3 {\rm :(}x_1 ,x_2 ,x_3 {\rm )} \to { (6,8)}, \\
 T_{3,2,1}^3 {\rm :(}x_1 ,x_2 ,x_3 {\rm )} \to { (4,8)}, \\
 \end{array}
\label{eqn7}
\end{eqnarray}

where $ j ,k$ and $ l \in \{1,2,3\}$  in any order.\\

\subsection*{The Peres gate $P$}

\begin{figure}
\begin{center}
\[
\Qcircuit @C=1.5em @R=0.5em @!R{
\lstick{x_1}       &\ctrl{1}           &\ctrl{1}           &\gate{C}\qwx[1]            &\targ\qwx[1]           &\gate{C}\qwx[1]        &\targ\qwx[1]       &\qw        &\rstick{y_1}\\
\lstick{x_2}     &\gate{C}\qwx[1]             &\targ  \qwx[1]     &\targ\qwx[1]       &\ctrl{1}\qwx[1]    &\targ\qwx[1]               &\gate{C}\qwx[1]                    &\qw        &\rstick{y_2}\\
\lstick{x_3}      &\targ       &\gate{C}               &\ctrl{0}           &\gate{C}           &\ctrl{0}       &\ctrl{0}               &\qw        &\rstick{y_3}\\
&P_{1,2,3} & P_{1,3,2}&P_{3,1,2}&P_{2,3,1}&P_{3,1,2}&P_{3,2,1}
}
\]
\caption{The possible $P$ gates over 3-bits reversible circuit.}
\label{fig4}
\end{center}
\end{figure}
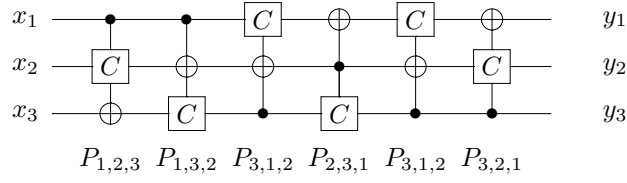

The $P$ gate is a 3-bits reversible gate. It combines the functions of $T^3$ gate and $C$ gate in
a single gate; and acts on an arbitrary 3-bits $x_j$, $x_k$ and
$x_l$. It applies the $C$ gate on one bit, taking the first bit as controller, and applies the $T^3$ gate on the other bit, taking the first and second bits as controller. For a $3\times 3$ bits circuit, there are six different $P$ gates, having a quantum cost of $four$. Eqn.\ref{eqn9} shows the functionality of the $P$ gate.  The circuit representation of the $P$ gate is shown in Fig. \ref{fig4} \cite{citation20, citation09}. 

\begin{eqnarray}
\begin{array}{l}
P^3_{j,k,l}:\\ y_j=x_j,\\ y_k=x_j \oplus x_k,\\y_l=x_l \oplus x_jx_k,\\
\\
 P_{123 } : (x_1 , x_2 , x_3 )  \to  (5, 7, 6, 8), \\
 P_{132}  : (x_1 , x_2 , x_3 )  \to  (5, 6, 7, 8), \\
 P_{213}  : (x_1 , x_2 , x_3 )  \to   (3, 7, 4, 8), \\
 P_{231}  : (x_1 , x_2 , x_3 )  \to  (3, 4, 7, 8), \\
 P_{312}  : (x_1 , x_2 , x_3 )  \to  (2, 6, 4, 8), \\
 P_{321}  : (x_1 , x_2 , x_3 )  \to  (2, 4, 6, 8). \\
 \end{array}
\label{eqn9}
\end{eqnarray}

\subsection*{The Fredkin gate $F$}

\begin{figure}
\begin{center}
\[
\Qcircuit @C=1.5em @R=1.0em @!R{
\lstick{x_1}         
&\qw  &\ctrl{1}   &\qw    &\qswap\qwx[1]  &\qw &\qswap\qwx[1] &\qw              &\rstick{y_1}\\
\lstick{x_2}         &\qw     &\qswap\qwx[1]&\qw  &\ctrl{1}   &\qw    &\qswap\qwx[1]  &\qw    &\rstick{y_2}\\
\lstick{x_3}        &\qw       &\qswap &\qw    &\qswap &\qw    &\ctrl{0}&\qw                &\rstick{y_3}\\
&&F_{1,2,3} &&F_{2,1,3} &&F_{3,2,1} &
}
\]
\caption{The possible $F$ gates over 3-bits reversible circuit.}
\label{fig5}
\end{center}
\end{figure}
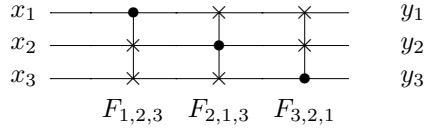

The $F$ gate is a 3-bits reversible gate. It is used to perform conditional swap on two of its
inputs if the third input is set to 1. For a $3\times 3$ bits circuits, there are three different $F$ gates, with quantum cost equals to
$five$ \cite{citation20}. Eqn.\ref{eqn8} shows the functionality of the
$F$ gate. The circuit representation of the $F$ gate is shown in Fig. \ref{fig5} \cite{citation09}.
\begin{eqnarray}
\begin{array}{l}
F^3_{j,k,l}: y_j=x_j,\\
\ \ \ \ \ \ \ \ \ x_j  = \left\{ {\begin{array}{*{20}c}
   {\begin{array}{*{20}c}
   1 & {y_k  = x_l ,y_l  = x_k }  \\
\end{array}}  \\
   {\begin{array}{*{20}c}
   0 & {y_k  = x_k ,y_l  = x_l }  \\
\end{array}}  \\
\end{array}} \right.

\\
\\
F_{1,2,3}^3 {\rm :(}x_1 ,x_2 ,x_3 {\rm )} \to { (6,7)}, \\
F_{2,1,3}^3 {\rm :(}x_1 ,x_2 ,x_3 {\rm )} \to { (4,7)}, \\
F_{3,2,1}^3 {\rm :(}x_1 ,x_2 ,x_3 {\rm )} \to { (4,6)}. \\
 \end{array}
\label{eqn8}
\end{eqnarray}

\subsection*{The $R$ gate}
The $R^n$ gate is $n$-bits universal reversible gate. It is first introduced in \cite{citation17n}. It combines the functionality of the $N$ gate, the $C$ gate and the $T$ gate. The $R^1$ gate is a 1-bit gate, which inverts the input bit unconditionally, there is one $R^1$ gate, it's quantum cost is $Zero$. Eqn.\ref{eqn15-10} shows the functionality of the $R^1$ gate \cite{citation17n}.\\

\begin{eqnarray}
\begin{array}{l}
R^1 _ i: y_i=x_i \oplus 1,\\
R_1^1:(x_1) \to (1,2).
 \end{array}
\label{eqn15-10}
\end{eqnarray}
 The $R^2$ gate is 2-bits gate, which makes one bit as a controller to flip the other bit, then it flips the controller bit is flipped unconditionally, there are two different $R^2$ gates, the quantum cost of the $R^2$ gate is $one$. Eqn.\ref{eqn15-11} shows the functionality of the $R^2$ gate \cite{citation17n}.\\

\begin{eqnarray}
\begin{array}{l}
R ^2_ {i,j}: y_i=x_i \oplus 1,\\
\ \ \ \ \ \ \ \ \   y_j=x_i \oplus x_j,\\
R_{1,2}^2:(x_1,x_2) \to (1,3,2,4),\\
R_{2,1}^2:(x_1,x_2) \to (1,2,3,4).
 \end{array}
\label{eqn15-11}
\end{eqnarray}

\begin{figure}
\begin{center}
\[
\Qcircuit @C=0.5em @R=0.5em @!R{
\lstick{x_1} &\gate{C_{2,1}}\qwx[1]                &\qw        &\gate{C_{3,1}}\qwx[1]       &\qw       &\gate{C_{2,1}}\qwx[1]          &\qw        &\gate{C_{3,1}}\qwx[1]       &\qw       &\gate{T^3N}\qwx[1]             &\qw        &\gate{T^3N}\qwx[1]                 &\qw        &\rstick{y_1}\\
\lstick{x_2} &\gate{T^3N}\qwx[1]               &\qw        &\gate{T^3N}\qwx[1]                 &\qw        &\gate{C_{3,2}}\qwx[1]              &\qw        &\gate{C_{1,2}}\qwx[1]              &\qw        &\gate{C_{3,2}}\qwx[1]              &\qw        &\gate{C_{1,2}}\qwx[1]      &\qw        &\rstick{y_2}\\
\lstick{x_3} &\gate{C_{1,3}}               &\qw &\gate{C_{2,3}}                &\qw        &\gate{T^3N}            &\qw        &\gate{T^3N}                &\qw        &\gate{C_{1,3}}                 &\qw        &\gate{C_{2,3}}                 &\qw        &\rstick{y_3}\\
& R_{1,2,3}^3 &&  R_{3,2,1}^3&&  R_{2,3,1}^3&&  R_{1,3,2}^3&&  R_{3,1,2}^3&&  R_{2,1,3}^3
}
\]
\caption{The six possible $R^ 3 $ gates for 3-bits reversible circuit.}
\label{fig12}
\end{center}
\end{figure}
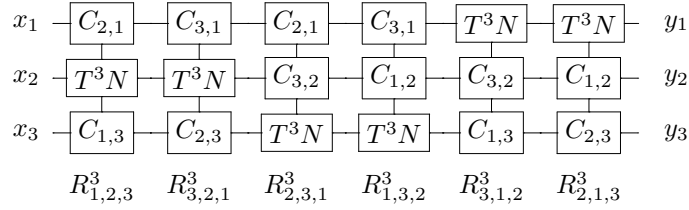

The $R^3$ gate is 3-bits reversible gate. The quantum cost for the $R^3$ gate is $four$. For 3-bits input/output circuit, there is six different $R^3$ gates as shown in Fig. \ref{fig12}. Eqn.\ref{eqn15} shows the functionality of the six different $R^3$ gates \cite{citation17n}.\\

\begin{eqnarray}
\begin{array}{l}
R_{j,k,l}^3: y_j=x_j \oplus x_k \oplus x_j . x_l\oplus 1 ,\\
\ \ \ \ \ \ \ \ \   y_k=x_k \oplus x_j . x_l\oplus 1, \\
\ \ \ \ \ \ \ \ \   y_l=x_l \oplus x_j,\\
\\
 R_{1,2,3}^3  :(x_1 ,x_2 ,x_3 ) \to (1 ,7 ,6 , 5 ,4 ,2 ,8 ,3), \\
 R_{3,2,1}^3 :(x_1 ,x_2 ,x_3 ) \to (1 ,4 ,6 ,2 ,7 ,5 ,8 ,3), \\
 R_{2,3,1}^3 :(x_1 ,x_2 ,x_3 ) \to (1 ,4 ,7 ,3 ,6 ,5 ,8 ,2),   \\
 R_{1,3,2}^3 :(x_1 ,x_2 ,x_3 ) \to (1 ,6 ,7 ,5 ,4 ,3 ,8 ,2), \\
 R_{3,1,2}^3 :(x_1 ,x_2 ,x_3 ) \to (1 ,6 ,4 ,2 ,7 ,3 ,8 ,5), \\
 R_{2,1,3}^3 :(x_1 ,x_2 ,x_3 ) \to (1,7 ,4 ,3 ,6 ,2 ,8 ,5), \\
 \end{array}
\label{eqn15}
\end{eqnarray}

For $n \ge 3$ $R^n$ combines the functionality of $N,C,T^3,T^4,...,T^n$, there are $n!$ different $R^n$ gates. In this paper we are going to build the proposed design using the $R^3$ gate.\\

\section{The Proposed Design}
This section introduces the two different proposed designs, which are half adder/subtractor and full adder/subtractor, designed using $R^3$ gate.\\

\subsection{The Proposed Design for half Adder/Subtractor}
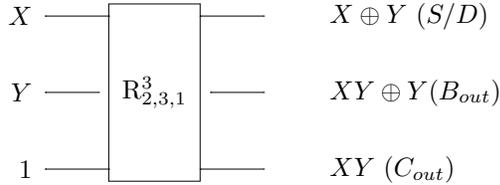
\begin{figure}[h]
\begin{center}
\[
\Qcircuit @C=2em @R=2em {
\lstick{X} & \multigate{2}{\mathcal{\rm R^3_{2,3,1}}} & \qw &\rstick{X \oplus Y\ (S/D)}\\
\lstick{Y} & \ghost{\mathcal{R^3_{2,3,1}}} & \qw  &\rstick{XY \oplus Y (B_{out})}\\
\lstick{1}& \ghost{\mathcal{\rm R^3_{2,3,1}}} & \qw &\rstick{XY\ (C_{out})}
}
\]
\caption{The proposed design of the half adder/subtractor.}
\label{fig14} 
\end{center}
\end{figure}

The half adder/subtractor is capable of adding or subtracting two bits $X$ and $Y$, realizing the operation $S=X+Y$, with $C_{out}$ as the carry out bit, after performing addition, and realizing the operation $D=X-Y$, with $B_{out}$ as the borrow out bit, after performing subtraction. The $R^3_{3,1,2}$ gate can be used by itself as a reversible half adder/subtractor. It takes as input 3-bits, the first two input bits are the two bits required to be added or subtracted $X$ and $Y$, the third bit is a constant bit of value one. It returns 3-bits as output, the first bit is the summation bit $S$ or the difference bit $D$, calculated by $X\oplus Y$, the second bit is the borrowing bit  $B_{out}$, calculating $\bar{X}Y$ and the third bit is the carry out $C_{out}$ bit, calculating $XY$. Fig. \ref{fig14} shows the circuit representation for half adder/subtractor, designed using $R^3_{3,1,2}$ gate.\\

The operation of the proposed half adder/subtractor can be expressed in the form of truth table, as shown in Table \ref{tbl1}. The quantum cost of the proposed design is $four$.\\

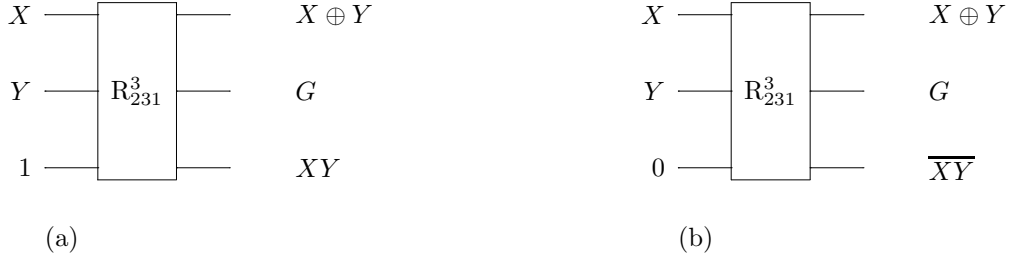
\begin{figure}[h]
\begin{tabular}{p{8cm}p{8cm}}
\Qcircuit @C=2em @R=2em {
\lstick{X}& \multigate{2}{\mathcal{\rm R^3_{231}
}} & \qw &\rstick{X \oplus Y}\\
\lstick{Y}& \ghost{\mathcal{\rm R^3_{231}
}} & \qw &\rstick{G}\\
\lstick{1}& \ghost{\mathcal{\rm R^3_{231}
}} & \qw &\rstick{XY}
}&
\Qcircuit @C=2em @R=2em {
\lstick{X} & \multigate{2}{\mathcal{\rm R^3_{231}}} & \qw &\rstick{X \oplus Y}\\
\lstick{Y} & \ghost{\mathcal{\rm R^3_{231}}} & \qw &\rstick{G}\\
\lstick{0}& \ghost{\mathcal{\rm R^3_{231}}} & \qw &\rstick{\overline{XY}}
}
\\\\ (a)&(b)
\end{tabular}
 \caption {Circuit representations for $R^3_{2,3,1}$ gate as logical operator where: (a) the constant bit is set to 1 to calculate $AND$ and $XOR$ operations, and  (b) the controller bit is set to 0 to calculate $NAND$ and $XOR$ operations.}.
\label{fig15}
\end{figure}

\begin{table}[]
\centering
\caption{The truth table for the proposed half Adder/subtractor.}
\label{tbl1}
\begin{tabular}{|c|c|c|c|c|c|}
\hline
$X$ & $Y$ & 1 & $S/D$ & $B_{out}$ & $C_{out}$ \\ \hline
0 & 0 & 1 & 0   & 0    & 0    \\ \hline
0 & 1 & 1 & 1   & 1    & 0    \\ \hline
1 & 0 & 1 & 1   & 0    & 0    \\ \hline
1 & 1 & 1 & 0   & 0    & 1    \\ \hline
\end{tabular}
\end{table}
The half adder/subtractor can be extended to perform different logical operation, by changing the order of the input bits and the value of the constant bit. Setting the constant bit to 1, allows the design to perform bit-wise $XOR$ and $AND$ operations, while setting the constant bit to 0, allows it perform bit-wise $XOR$ and $NAND$ operation, as shown in Fig \ref{fig15}, where $G$ is the garbage bit. Setting the value of the input $Y$ bit to 1, allows the design to perform bit-wise complement and copy of basis for the input bit $X$, as shown in Fig \ref{fig15-1}. Changing the order of the input bits, in which the first input bit is set to 1 as a constant bit, allows the design to perform $XNOR$ operation as shown in Fig \ref{fig15-2}.\\
%
\begin{figure}[h]
\begin{tabular}{p{8cm}}
\Qcircuit @C=2em @R=2em {
\lstick{X}& \multigate{2}{\mathcal{\rm R^3_{231}}} & \qw &\rstick{\bar{X}}\\
\lstick{1}& \ghost{\mathcal{\rm R^3_{231}}} & \qw &\rstick{G}\\
\lstick{1}& \ghost{\mathcal{\rm R^3_{231}}} & \qw &\rstick{X}
}
\end{tabular}
 \caption {Circuit representations for $R^3_{2,3,1}$ gate as logical operator where the constant bit is set to 1 and one of the other two bits is set to 1, to calculate $NOT$ and copy of basis operations.}.
\label{fig15-1}
\end{figure}
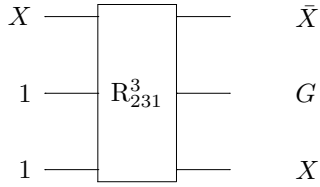
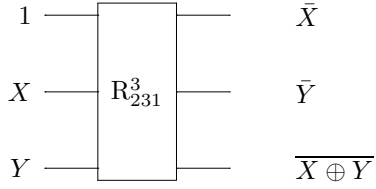
\begin{figure}[h]
\begin{tabular}{p{8cm}}
\Qcircuit @C=2em @R=2em {
\lstick{1}& \multigate{2}{\mathcal{\rm R^3_{231}}} & \qw &\rstick{\bar{X}}\\
\lstick{X}& \ghost{\mathcal{\rm R^3_{231}}} & \qw &\rstick{\bar{Y}}\\
\lstick{Y}& \ghost{\mathcal{\rm R^3_{231}}} & \qw &\rstick{\overline{X \oplus Y}}
}
\end{tabular}
 \caption {Circuit representations for $R^3_{2,3,1}$ gate as logical operator, to calculate $XNOR$ operations.}.
\label{fig15-2}
\end{figure}

\subsection{The Proposed Design for full Adder/Subtractor}
A half adder/subtractor is not very useful on its own. In addition, a third bit is needed to add the carry-in to the two addend, while in subtraction, the borrow-in is required to be added to the subtrahend, thus a full adder/subtractor is required. It realizes the two operations $S=X+Y+Z$ and $D=X-Y-Z$, where $X$ and $Y$ are the two input bits required to be added or subtracted, and $Z$ is either the carry-in bit during addition or the borrow-in bit during subtraction.\\

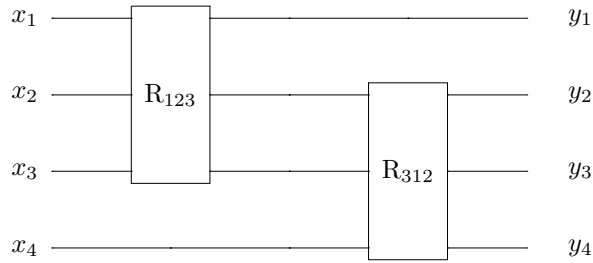
\begin{figure}[h]
\begin{center}
\[
\Qcircuit @C=3em @R=2em {
\lstick{x_1}& \multigate{2}{\mathcal{\rm  R_{123}}} & \qw & \qw & \qw &\lstick{y_1}  \\
\lstick{x_2}& \ghost{\mathcal{\rm  R_{123}}} & \qw & \multigate{2}{\mathcal{\rm  R_{312}}} & \qw  &\lstick{y_2}\\
\lstick{x_3}& \ghost{\mathcal{\rm  R_{123}}} & \qw & \ghost{\mathcal{\rm  R_{312}}} & \qw & \lstick{y_3}\\
\lstick{x_4}& \qw & \qw  & \ghost{\mathcal{\rm  R_{312}}} & \qw & \lstick{y_4}
}
\]
\caption{The proposed reversible Full Adder/subtractor using $R$ gate library.}
\label{fig16} 
\end{center}
\end{figure}

The proposed design of the full adder/subtractor consists of two $R^3$ gates, which are: $R^3_{1,2,3}$ and $R^3_{3,1,2}$. The two inputs $X$ and $Y$ with a constant bit is passed to $R^3_{1,2,3}$ gate, The first output is a garbage bit, the other two outputs are passed as input to $R^3_{3,1,2}$ with $Z$ bit, which is either $C_{in}$ during addition and $B_{in}$ during subtraction. The output of this gate is 3-bits, which are: carry-out ($C_{out}$) at $y_2$, borrow-out ($B_{out}$) at $y_4$ and sum ($S$) or difference ($D$) at $y_3$, as shown in Fig. \ref{fig16}.\\
%
%

%
The proposed design can work as a reversible full adder, if the value of the constant bit at $x_2$ is set to 0, while it can work as a reversible full subtractor, if the value of $Y$ is entered instead of the constant bit at $x_2$. Eqn.\ref{eqn16} shows the equations of the output bits. Table \ref{tbl2} shows the truth table of the proposed reversible full adder, and Table \ref{tbl3} shows the truth table of the proposed reversible  full subtractor. The quantum cost of the proposed full adder/subtractor is $eight$. The following examples shows in steps how addition and subtraction are calculated using the proposed full adder/subtractor.  \\
\newline

\begin{exmp}

suppose we want to add the values(1+1+1=3) then $X=1$, $Y=1$ and $C_{in}=1$, the inputs to $R_{123}$ are $x_1=1, x_2=0$ and $x_3=1$, giving outputs $y_1=0$, 1 and 0. 1 and 0 with $x_4=0$ are passed as input to $R_{312}$, giving outputs $y_2=1, y_3=0$ and $y_4=0$. Since $y_3$ is the Difference bit and $y_4$ is the $B_{out}$ bit, and they have the values 0,0 respectively, then the output is 0.\\ 

\end{exmp}

\begin{exmp} 

suppose we want to subtract the values(1-1=0) then $X=1$, $Y=1$ and $B_{in}=0$, the inputs to $R_{123}$ are $x_1=1, x_2=1$ and $x_3=1$, giving outputs $y_1=0$, 0 and 0. 0 and 0 with $x_4=1$ are passed as input to $R_{312}$, giving outputs $y_2=1, y_3=1$ and $y_4=0$. Since $y_2$ is the $C_out$ bit and $y_3$ is the summation bit, and they have the values 1,1 respectively, then the output is 3.\\ 

\end{exmp}

\begin{eqnarray}
\begin{array}{l}
S= X \oplus Y \oplus Z,\\
C_{out}=XY \oplus XZ \oplus YZ.\\
D=X \oplus Y \oplus Z.\\
B_{out}=\bar{X}(Y \oplus Z) \oplus YZ,
 \end{array}
\label{eqn16}
\end{eqnarray}

\noindent where $X$ and $Y$ are the input bits, which are needed to be added or subtracted, $Z$ is either the carry-in bit, when addition is applied or the borrow-in bit, when subtraction is applied. $S$ is the summation bit, $D$ is the difference bit, $C_{out}$ is the carry out bit, $B_{out}$ is the borrow out bit.\\

\begin{table}[]
\centering
\caption{The truth table of the proposed full adder.}
\label{tbl2}
\begin{tabular}{|c|c|c|c|c|}
\hline
\textbf{$X$}  & \textbf{$Y$} & \textbf{$Z$} & \textbf{$S$} & \textbf{$C_{out}$} \\ \hline
0                   & 0          & 0          & 0          & 0             \\ \hline
0                   & 0          & 1          & 1          & 0             \\ \hline
0                   & 1          & 0          & 1          & 0             \\ \hline
0                   & 1          & 1          & 0          & 1             \\ \hline
1                   & 0          & 0          & 1          & 0             \\ \hline
1                   & 0          & 1          & 0          & 1             \\ \hline
1                   & 1          & 0          & 0          & 1             \\ \hline
1                   & 1          & 1          & 1          & 1             \\ \hline
\end{tabular}
\end{table}

\begin{table}[]
\centering
\caption{The truth table of the proposed full subtractor.}
\label{tbl3}
\begin{tabular}{|c|c|c|c|c|c|}
\hline
\textbf{$X$}  & \textbf{$Y$} & \textbf{$Z$} & \textbf{$D$} & \textbf{$B_{out}$} \\ \hline
0                  & 0          & 0          & 0          & 0             \\ \hline
0                  & 0          & 1          & 1          & 1             \\ \hline
0                  & 1          & 0          & 1          & 1             \\ \hline
0                  & 1          & 1          & 0          & 1             \\ \hline
1                  & 0          & 0          & 1          & 0             \\ \hline
1                  & 0          & 1          & 0          & 0             \\ \hline
1                  & 1          & 0          & 0          & 0             \\ \hline
1                   & 1          & 1          & 1          & 1             \\ \hline
\end{tabular}
\end{table}

\begin{table}[]
\centering
\caption{The effect of the constant inputs on the result of the 1-bit ALU proposed design based on Fig.\ref{fig16}}
\label{tbl3-1}
\begin{center}
\begin{tabular}{|l|l|l|l||l|l| l|l||l|}
\hline
\textbf{$x_1$} & \textbf{$x_2$} & \textbf{$x_3$} & \textbf{$x_4$} & \textbf{$y_1$} & \textbf{$y_2$} & \textbf{$y_3$}          & \textbf{$y_4$} & \textbf{Result} \\[2ex] \hline
$X$            & 0              & $Y$            & $C_{in}$       & $G$            & $C_{out}$      & $Sum$                   & $G$            &  ADD             \\[2ex] \hline
$X$            & $Y$            & $Y$            & $B_{in}$       & $G$            & $G$            & $Diff$                  & $B_{out}$      &  SUB             \\[2ex] \hline
$X$            & 0              & $Y$            & -              & $G$            & $XY$           & $G$                     & $G$            & AND             \\[2ex] \hline
$X$            & 0              & $Y$            & 0              & $G$            & $G$            & $X \oplus Y$            & $G$            & XOR             \\[2ex] \hline
$X$            & 0              & $Y$            & 1              & $G$            & $G$            & $\overline{X \oplus Y}$ & $G$            & XNOR            \\[2ex] \hline
$X$            & 0              & 1              & 0              & $G$            & $G$            & $\overline{X} $       & $G$            & NOT             \\[2ex] \hline
\end{tabular}
\end{center}
\end{table}

\subsection{Experimental Results}
The proposed designs are compared with the designs proposed by others, according to the number of gates used to build the design, the number of constant bits, the number of garbage bits and the quantum cost of the design.\\

The comparison between the proposed half adder/subtractor design and the designs proposed by others, is shown in Table \ref{tbl5}. It shows that the number of garbage bits achieved by the proposed design is 0, which is the best number of garbage bits achieved by the designs proposed by others. The number of gates used to build the proposed design is 1, which is similar to the best number of gates used to build designs proposed by others. The quantum cost of the proposed design is $four$, similar to the best quantum cost achieved by \cite{citation18n}. The number of constant bits is 1, similar to the smallest number of constant bits achieved by \cite{citation18n,new2}. As shown in Table \ref{tbl5} the proposed design shows the best results, compared with the other designs.\\

\begin{table}[]
\centering
\caption{Comparing the different designs of half adders and subtractors.}
\label{tbl5}
\begin{tabular}{|l|l|l|l|l|l|l|}
\hline
\textbf{}        & \textbf{\begin{tabular}[c]{@{}l@{}}Function\end{tabular}} & \textbf{Gates used}                                                 & \textbf{\begin{tabular}[c]{@{}l@{}}No.\\ Gates\end{tabular}} & \textbf{\begin{tabular}[c]{@{}l@{}}No.\\ Garbage \\ bits\end{tabular}} & \textbf{\begin{tabular}[c]{@{}l@{}}No.\\ Conatant\\ bits\end{tabular}} & \textbf{\begin{tabular}[c]{@{}l@{}}Quantum\\ Cost\end{tabular}} \\ \hline
\begin{tabular}[c]{@{}l@{}} Proposed \\ Design \end{tabular} &  ADD/SUB                                                     & One $R^3_{3,1,2}$                                                            & 1                                                            & 0                                                                      & 1                                                                      & 4                                                               \\ \hline
{}\cite{citation12n}{}         & ADD/SUB                                                      & \begin{tabular}[c]{@{}l@{}}2 $C$ gates+\\ 2 $Mux$ gates\end{tabular} & 4                                                            & 3                                                                      & 3                                                                      & 8                                                               \\ \hline
{}\cite{citation11n6}{}        &  ADD/SUB                                                      & One $NR$ gate                                                         & 1                                                            & 1                                                                      & 2                                                                      & 7                                                               \\ \hline
{}\cite{citation10n}{}         & ADD/SUB                                                      & One $MOG$ gate                                                        & 1                                                            & 2                                                                      & 2                                                                      & 11                                                              \\ \hline
{}\cite{citation18n}{}        &SUB                                                 & One $TR$ gate                                                         & 1                                                            & 1                                                                      & 1                                                                      & 4                                                               \\ \hline
{}\cite{new2}{}        &      ADD                                            &\begin{tabular}[c]{@{}l@{}}One $T$ gate+\\ one controlled\\ controlled Z(-1) \\gate+ one\\ $C$ gate \end{tabular}                                                          & 3                                                            & 1                                                                      & 1                                                                      &17                                                               \\ \hline
\end{tabular}
\end{table}

The comparison between the proposed design used to perform logical operations, and the designs proposed by others is shown in Table \ref{tbl6}.The first design performs $AND$, $XOR$ and $NAND$ operations. The number of gates used to build the design is 1, similar to the best number of gates used to build the designs proposed by others. The number of garbage bits is 1, which is also similar to the best number of garbage bits achieved by \cite{citation07n,citation17n-1}. The number of constant bits used in the proposed design is 1, similar to the best number of constant bits used by the best designs proposed by others. The quantum cost of the proposed design is $four$, better than all the designs proposed by others.\\

The $NOT$ and copy of basis operations can be achieved by the second design. By comparing it with the other proposed designs, it is found that, the number of gates used to build the proposed design is 1, similar to the best number of gates used to build the designs proposed by others. The number of garbage bits in the proposed design is 1, similar to the best number of garbage bits achieved by \cite{citation07n,citation17n-1}. The quantum cost of the proposed design is $four$, better than the quantum cost of the other designs proposed by others.\\

The $XNOR$ and the $NOT$ operations can be obtained by the third proposed design. By comparing the four criteria points in the proposed design with the designs proposed by others, it is found that, the number of gates used to build the proposed design is 1, similar to the best number of gates used in the other designs. The number of garbage bits in the proposed design is 0, which is less than all the other designs proposed by others. The number of constant bits in the proposed design is 1, similar to the number of constant bits, used in the best designs proposed by others. The quantum cost of the proposed design is $four$, which is better than the quantum cost of the designs proposed by others.\\ 

\begin{table}[]
\centering
\caption{Comparing the $R$ gate the other designs used to perform logical operation}
\label{tbl6}
\begin{tabular}{|l|l|l|l|l|l|l|}
\hline
\textbf{}              & \textbf{\begin{tabular}[c]{@{}l@{}}Logic\\ operation\end{tabular}} & \textbf{Gates used} & \textbf{\begin{tabular}[c]{@{}l@{}}No.\\ Gates\end{tabular}} & \textbf{\begin{tabular}[c]{@{}l@{}}No.\\ Garbage \\ bits\end{tabular}} & \textbf{\begin{tabular}[c]{@{}l@{}}No.\\ Conatant\\ bits\end{tabular}} & \textbf{\begin{tabular}[c]{@{}l@{}}Quantum\\ Cost\end{tabular}} \\ \hline
Proposed Design        & \begin{tabular}[c]{@{}l@{}}AND+ XOR\\ + NAND\end{tabular}          & One $R^3_{3,1,2}$               & 1                                                            & 1                                                                      & 1                                                                      & 4                                                               \\ \hline
Proposed Design        & Copy + NOT                                                         & One $R^3_{3,1,2}$               & 1                                                            & 1                                                                      & 2                                                                      & 4                                                               \\ \hline
Proposed Design        & XNOR+NOT           & One $R^3_{3,1,2}$               & 1                                                            & 0                                                                      & 1                                                                      & 4                                                               \\ \hline
{}\cite{citation17n-1}{} & OR                                                                 & one $RG1$ gate        & 1                                                            & 2                                                                      & 1                                                                      & 5                                                               \\ \hline
{}\cite{citation17n-1}{} & AND                                                                & One $RG1$ gate        & 1                                                            & 2                                                                      & 1                                                                      & 5                                                               \\ \hline
{}\cite{citation17n-1}{} & XOR                                                                & One $RG1$ gate        & 1                                                            & 2                                                                      & 1                                                                      & 5                                                               \\ \hline
{}\cite{citation17n-1}{} & NOT + copy                                                         & One RG1 gate        & 1                                                            & 1                                                                      & 2                                                                      & 5                                                               \\ \hline
{}\cite{citation17n-1}{} & NOR                                                                & One $RG2$ gate        & 1                                                            & 2                                                                      & 1                                                                      & 5                                                               \\ \hline
{}\cite{citation07n}{} & OR+AND                                                             & One $F$ gate          & 1                                                            & 1                                                                      & 1                                                                      & 5                                                               \\ \hline
{}\cite{citation07n}{}                    & XOR                                                                & 2 $F$ gate            & 2                                                            & 3                                                                      & 2                                                                      & 10                                                              \\ \hline
{}\cite{new3}{}                    &\begin{tabular}[c]{@{}l@{}}OR+XOR+\\NOR+XNOR\end{tabular}                                                                & one $MRG$ gate            & 1                                                            & 2                                                                      & 2                                                                      & 6 \\ \hline
{}\cite{citation28-1}{}                    &\begin{tabular}[c]{@{}l@{}}NOT+XOR\\+NOR\end{tabular}                                                                 & one $TSG$ gate            & 1                                                            & 3                                                                      & 1                                                                      & 6 \\ \hline
{}\cite{citation28-1}{}                    &\begin{tabular}[c]{@{}l@{}}XOR+NAND+\\XNOR+OR \end{tabular}                                                             & one $HNG$ gate            & 1                                                            & 2                                                                      & 2                                                                      & -\\ \hline
{}\cite{citation28-1}{}                    &\begin{tabular}[c]{@{}l@{}}NOR+AND\\+NOT \end{tabular}                                                               & one $HNG$ gate            & 1                                                            & 3                                                                      & 2                                                                      & - \\ \hline
\end{tabular}
\end{table}

The comparison between the proposed full adder/subtractor and the designs proposed by others is shown in Table \ref{tbl4}. Some of the comparable designs in Table \ref{tbl4} are full adder/subtractor, such as: \cite{citation10n,citation11n6,citation14n,citation12n,new3,new4,new1}{}, other are full adders only, such as: \cite{citation01n,citation03n,citation04n,citation07n,citation15n,new5,new2}{}, while in \cite{citation18n,citation16n-1}{}  full subtractors are proposed.\\

The proposed full adder/subtractor is composed of two gates, which is the second best number of gates compared by the other designs, since in \cite{citation10n,citation16n-1} only one gate is used. The number of garbage bits in the proposed design is 1, similar to the number of garbage bits in \cite{citation11n6}, which is the best number of garbage bits achieved by the designs proposed by others. The number of constant bits in the proposed design is 1, similar to the second best number of constant bits used in the designs proposed by others, since the best number of constant bit is 0 proposed by \cite{citation10n}{}. The quantum cost of the proposed design is 8, which is similar to the best quantum cost achieved by the full adder/subtractor described in 
\cite{citation04n}. The quantum cost of the proposed design is greater than the best quantum cost of the full subtractor designed by \cite{citation18n}{}. The delay of the proposed design is 8, which is similar to the full adder/subtractor design in \cite{citation11n6}, much less than the delay of the full adder/subtractor design in \cite{new3}. The delay of the proposed design is higher than the delay of the designs in \cite{citation18n,new5}, but these two designs are either full subtractor or full adder respectively. \\

The proposed design can be used as 1-bit ALU as it can perform Addition (ADD), Subtraction(SUB) and different logical operations such as: AND, XOR, XNOR and NOT, depending of the on the values of the constant bits, as shown in Table \ref{tbl3-1}. By comparing the proposed design with the deign described in \cite{new3}, the proposed design has less quantum cost and less number of garbage bits.\\

Comparing the proposed design with the previously introduced designs, shows that the proposed design achieves better performance compared with the existing designs in terms of the number of gates used, the number of constant bits, the number of garbage bits and the quantum cost.  \\

\begin{table}[]
\centering
\caption{Comparing the different designs of full adders and subtractors.}
\label{tbl4}
\begin{tabular}{|l|l|l|l|l|l|l|l|}
\hline
\multicolumn{1}{|c|}{\textbf{}} & 
Function& \multicolumn{1}{c|}{\textbf{Gates used}}                                            & \multicolumn{1}{c|}{\textbf{\begin{tabular}[c]{@{}c@{}}No. \\ gates\end{tabular}}} & \multicolumn{1}{c|}{\textbf{\begin{tabular}[c]{@{}c@{}}Garbage \\ bits\end{tabular}}} & \multicolumn{1}{c|}{\textbf{\begin{tabular}[c]{@{}c@{}}Constant \\ bits\end{tabular}}} & \multicolumn{1}{c|}{\textbf{\begin{tabular}[c]{@{}c@{}}Quantum \\ cost\end{tabular}}} &Delay\\ \hline
\begin{tabular}[c]{@{}l@{}}Proposed \\ Design\end{tabular}  & ADD/SUB                  & 2 $R^3$ gates                                                                          & 2                  & 1                     & 1                      & 8  &8                   \\ \hline
{}\cite{citation10n}{}        & ADD/SUB                  & One $MOG$ gate                                                                        & 1                  & 2                     & 0                      & 11          &-          \\ \hline
{}\cite{citation11n6}{}        & ADD/SUB                   & 2 $NR$ gate                                                                           & 2                  & 1                     & 2                      & 14          &8          \\ \hline
{}\cite{citation12n}{}         & ADD/SUB                    & \begin{tabular}[c]{@{}l@{}}2 $Mux$ gates + \\ One $TR$ gate + \\ 5 $F$ gates\end{tabular} & 8                  & 5                     & 3                      & 19   &-                 \\ \hline
{}\cite{citation14n}{} Design1 & ADD/SUB                   & \begin{tabular}[c]{@{}l@{}}5 $C$ gates+\\ 2 $F$ gates +\\ one $TR$ gate\end{tabular}   & 8                  & 5                     & 3                      & 21   &-                 \\ \hline
{}\cite{citation14n}{} Design2 & ADD/SUB                    & \begin{tabular}[c]{@{}l@{}}2 $TR$ gates+\\ 2 $C$ gates\end{tabular}                  & 4                  & 3                     & 1                      & 14      &-              \\ \hline
{}\cite{citation14n} Design3    & ADD/SUB                   & \begin{tabular}[c]{@{}l@{}}2 $P$ gates+\\ 2 $C$ gates\end{tabular}                   & 4                  & 3                     & 1                      & 10      &-              \\ \hline
{}\cite{new3}{}         &   ADD/SUB                   &\begin{tabular}[c]{@{}l@{}}2 $C$ gates +\\one $HNG$ gate+\\2$F$ gates+\\one $PAOG$                                                                         gate  \end{tabular}    & 6                  & 4                     & 5                      & 24          &20           \\ \hline
{}\cite{new1}{}         &ADD/SUB                   &\begin{tabular}[c]{@{}l@{}}one $C$ gate + \\2$QR$ gates+\\one $NOT$ gate \end{tabular}   & 4                & 2                     & 1                     &-&- \\ \hline
{}\cite{citation15n}{}         & ADD                     & \begin{tabular}[c]{@{}l@{}}One $NG$ gate+\\ One $T^3$ gate+\\ One C\end{tabular}      & 3                  & 2                     & 1                      & 10  &-                  \\ \hline
{}\cite{citation04n}{}         & ADD                  & 2 P gates                                                                           & 2                  & 2                     & 2                      & 8                   &-  \\ \hline
{}\cite{citation07n}{}          & ADD                    & 4 $F$ gates                                                                           & 4                  & 4                     & 2                      & 20             &-       \\ \hline
{}\cite{citation03n}{}          & ADD                     & 2 $T^3$ + 2 $C$                                                                       & 4                  & 2                     & 2                      & 12         &-           \\ \hline
{}\cite{citation01n}{}          & ADD                     & \begin{tabular}[c]{@{}l@{}}One $RG1$\_1+\\ One $RG2$\_1\end{tabular}                    & 2                  & 2                     & 1                      & 10   &-                 \\ \hline
{}\cite{new5}{}         &   ADD                   &\begin{tabular}[c]{@{}l@{}}2 $C$ gates+\\3 $u$ gates + one\\ $v$ gate \end{tabular}  & 6                 & 2                     & 1                     & 6 &4\\ \hline
{}\cite{citation16n-1}{}         &   ADD                   &one $DPG$ gate  & 1                & 2                     & 1                     & 6&- \\ \hline
{}\cite{new2}{}         &ADD                   &\begin{tabular}[c]{@{}l@{}}2 $T$ gate + \\2$C$ gate + \\2 controlled \\controlled Z(-1)\\ gates \end{tabular}
& 6               & 2                     & 1                     &34&- \\ \hline
{}\cite{citation18n}{}         &    SUB                 & 2 $TR$ gates                                                                          & 2                  & 2                     & 1                      & 6               &4      \\ \hline
{}\cite{citation16n-1}{}         &     SUB                 &\begin{tabular}[c]{@{}l@{}}one $F$ gate+ \\2 $TSG$ gate \end{tabular}  & 3                & 6                     & 2                     &9&- \\ \hline
\end{tabular}
\end{table}
\section{Conclusion}
\label{sec:Conclusion}
The computer processor is the basic component of the computer. The most important part of the computer processor is the arithmetic logic unit. It performs binary addition, subtraction, multiplication and division. Addition and subtraction are the main operations in the arithmetic logical unit, since division and multiplication can be calculated using repeated subtraction and repeated division. Many applications uses the adder/subtractor, such as Arithmetic and logical unit(ALU), Program status word (PSW), Calculators, Embedded system, seven segment display etc. \\

In this paper, We proposed two new designs: quantum half adder/subtractor and quantum  full Adder/ Subtractor using $R^3$ gate. we have shown that these two designs can be used to perform the logical operations, such as: AND, XOR, NAND, NOT and XNOR. The proposed designs are compared with the other previously proposed designs, according to the number of gates used to build the design, the quantum cost, the number of constant bits and the number of garbage bits.\\

The proposed half adder/subtractor is synthesized using $R^3_{231}$ gate and can be used to perform logical operations. It is compared with the other previous designs and it has the minimum number of gates, the minimum number of garbage bits, the minimum number of constant bits and the minimum quantum cost.\\

The proposed full adder/subtractor is synthesized using $R^3_{123}$ and $R^3_{312}$ gates, it can work as 1-bit ALU. It can also extend to work on any number of bits, by adding multiplexer to link each full adder/subtractor.\\

It is shown that the proposed full adder/subtractor is build using 2 gates which is the second minimum number of gates, it has one garbage bits which is equal to the minimum number of garbage bits, it has one constant bit, which is equal to the second minimum number of constant bits and it has the minimum quantum cost compared with the other full adder/subtractor. Consequently it will improve the efficiency of the arithmetic logical unit.\\



\end{document}